# Quantifying a Firm's AI Engagement: Constructing Objective, Data-Driven, AI Stock Indices Using SEC 10-K Filings

## Lennart Ante[ac]*

*a Constructor University, Germany*
*c Blockchain Research Lab, Germany*
*ORCID: 0000-0002-5088-7127*

## Aman Saggu[bc]**

*b Business Administration Division, Mahidol University International College, Mahidol University*
*c Blockchain Research Lab*
*ORCID: 0000-0003-2784-4353*

This Version: January 1, 2024

**Abstract:** Following an analysis of existing AI-related exchange-traded funds (ETFs), we reveal the selection criteria for determining which stocks qualify as AI-related are often opaque and rely on vague phrases and subjective judgments. This paper proposes a new, objective, data-driven approach using natural language processing (NLP) techniques to classify AI stocks by analyzing annual 10-K filings from 3,395 NASDAQ-listed firms between 2011 and 2023. This analysis quantifies each company's engagement with AI through binary indicators and weighted AI scores based on the frequency and context of AI-related terms. Using these metrics, we construct four AI stock indices—the Equally Weighted AI Index (AII), the Size-Weighted AI Index (SAII), and two Time-Discounted AI Indices (TAII05 and TAII5X)—offering different perspectives on AI investment. We validate our methodology through an event study on the launch of OpenAI's ChatGPT, demonstrating that companies with higher AI engagement saw significantly greater positive abnormal returns, with analyses supporting the predictive power of our AI measures. Our indices perform on par with or surpass 14 existing AI-themed ETFs and the Nasdaq Composite Index in risk-return profiles, market responsiveness, and overall performance, achieving higher average daily returns and risk-adjusted metrics without increased volatility. These results suggest our NLP-based approach offers a reliable, market-responsive, and cost-effective alternative to existing AI-related ETF products. Our innovative methodology can also guide investors, asset managers, and policymakers in using corporate data to construct other thematic portfolios, contributing to a more transparent, data-driven, and competitive approach.

*Corresponding author: Dr. Lennart Ante, B.S., M.S., Ph.D., Chief Executive Officer, Blockchain Research Lab, gGmbH, Weidestraße 120b, 22083 Hamburg, Germany. E-mail address: ante@blockchainresearchlab.org.
**Corresponding author: Aman Saggu, Business Administration Division, Mahidol University International College, 999 Phutthamonthon Sai 4 Rd, Salaya, Phutthamonthon District, Nakhon Pathom 73170, Thailand, Telephone: 0066 27005000, Fax: 0066 24415091, Email address: asaggu26@gmail.com; aman.sag@mahidol.edu.

# 1. Introduction

The launch of OpenAI's ChatGPT in November 2022 catalyzed a global surge in corporate investment and public interest in artificial intelligence (AI). Companies across various sectors significantly increased their investments in AI technologies; however, some have exaggerated their involvement, leading to regulatory scrutiny. For instance, the SEC charged two investment advisers for making misleading AI-related disclosures, a practice termed "AI washing" (SEC, 2024). Concurrently, there has been a proliferation of AI-related Exchange-Traded Funds (ETFs), raising questions about the expertise of asset managers in the complex domain of AI technologies. We find that the composition of these AI ETFs often reflects subjective decision-making rather than a reliance on robust, objective, quantifiable metrics. This raises a critical inquiry: How equipped are traditional asset managers—individuals who may not necessarily have a background in technology—to assess the AI-relatedness of a company accurately?

Defining what constitutes an "AI stock" or an "AI-themed" investment is a complex challenge that lies at the heart of this research. We aim to develop a scholarly framework to understand and categorize such investments by first clarifying AI's role within the investment arena: Is AI utilized as a tool for investment management, or do the investments represent stakes in AI technology companies? In this study, we concentrate on stocks and indices that provide investors with opportunities to engage with and profit from AI technology itself. This focus raises a pivotal question: How can a stock or index be accurately identified as AI-centric? For example, consider Nvidia—a company crucial to the AI industry due to its production of chips essential for AI computing. Yet, Nvidia's influence extends beyond AI, impacting industries such as gaming and cryptocurrency mining. This underscores a fundamental insight: stocks or indices are rarely exclusively tied to a single technology like AI. Instead, they are often strategically positioned to maximize AI exposure while acknowledging their multi-sectoral influence.

A preliminary examination of thematic indices focused on AI reveals that the selection criteria for investments are often opaque, relying on vague phrases such as "*companies that potentially stand to benefit from increased adoption and utilization of […] artificial intelligence*" (Global X, 2023a) or "*companies that could benefit from the long-term growth and innovation in […] artificial intelligence*" (iShares, 2023). These subjective statements hinder the development of a clear, general concept of "AI stocks." To address the challenge of objectively identifying AI-centric investments, our study develops a data-driven methodology for classifying AI stocks based on publicly available corporate disclosures. We leverage natural language processing (NLP) techniques to analyze annual 10-K filings from 3,395 NASDAQ-listed firms spanning 2010 to 2022, aiming to quantify each company's engagement with AI technologies. Our approach involves creating both a binary indicator of AI mention and a weighted AI score that reflects the frequency of AI-related terms in these disclosures. Specifically, our methodology begins with the extraction and preprocessing of textual data from the Securities and Exchange Commission's (SEC) Electronic Data Gathering, Analysis, and Retrieval (EDGAR) database. We identify specific AI-related keywords through bibliometric analysis, focusing on terms such as "artificial intelligence" and "AI." By calculating term frequencies and applying the term frequency-inverse document frequency (TF-IDF) weighting scheme, we obtain a measure of each firm's emphasis on AI in their filings (see Section 3.2). This allows us to objectively assess the degree to which companies are involved in AI based on their own disclosures. Using these AI engagement metrics, we construct four AI stock indices: the Equally Weighted AI Index (AII), the Size-Weighted AI Index (SAII), and two Time-Discounted AI Indices (TAII05 and TAII5X). These indices differ in how they weigh companies based on their AI engagement and account for historical disclosures, providing alternative perspectives on AI investment opportunities. We then compare the performance of these indices against 14 existing AI-themed ETFs, analyzing their risk-return profiles, market reactions to the ChatGPT launch, and overall investment performance.

To validate our AI classification methodology, we conduct an event study centered on the market reaction to the launch of ChatGPT. We hypothesize that companies with a higher degree of AI engagement, as indicated by our metrics, would exhibit more significant positive abnormal returns following this pivotal AI-related event. Our findings support this hypothesis, demonstrating that our objective measures effectively capture market perceptions of AI involvement. Regression analyses further support the validity of our metrics, revealing a significant predictive power of our AI measures and the cumulative average abnormal returns (CAARs) observed. Comparative analysis shows that our objectively constructed AI indices perform on par with, or even surpass, existing AI-themed ETFs in terms of market response and investment performance. This suggests that our methodology offers a reliable and market-responsive approach to creating AI stock indices, providing investors with a viable alternative or complement to existing AI-themed investment products. Such an approach could enhance competition among index providers and potentially reduce ETF expense ratios (An et al., 2023), benefiting investors seeking exposure to the AI sector.

In summary, our study presents a coherent and logical framework for objectively identifying and evaluating AI-focused investments. By utilizing NLP-based analysis of corporate disclosures, we create data-driven AI engagement metrics that inform the construction of competitive AI stock indices. Our methodology addresses the opacity and subjectivity found in existing AI-themed ETFs and offers empirical support through market reaction analyses. This contributes to the literature on thematic investing and offers practical implications for investors and asset managers in the rapidly evolving AI investment landscape.

Our study makes meaningful contributions to the existing literature on financial markets and corporate disclosures, with a particular focus on AI technology. We address a notable gap in current research, contributing to the development of measures on the basis of company filings, such as 10-Ks (Andreou et al. 2020). Building upon the foundational work of Beaver (1968) and Griffin (2003) on how corporate information influences stock valuations, our results enhance the understanding of market reactions to technology disclosures in corporate filings (Aharon et al., 2022; Cahill et al., 2020; Cheng et al., 2019). Specifically, by analyzing the market reaction to the launch of OpenAI's ChatGPT as a validation measure for AI-focused stocks, we contribute to research on technology launches and their impact on financial markets, particularly in the context of AI advancements (Ante and Demir, 2024; Saggu and Ante, 2023). Furthermore, our comparative analysis of our data-driven AI stock indices and existing AI-themed ETFs uncovers the relative (in-)effectiveness of the latter. The findings suggest that asset managers and index providers should incorporate quantitative, objective metrics when selecting assets for inclusion, contributing to the literature on ETF selection criteria. Notably, our results indicate that index providers may be overcharging for their services—potentially accounting for over one-third of expense ratios (An et al., 2023)—when a straightforward NLP-based approach can yield comparable or even superior results.

By providing both methodology and empirical evidence on the identification of AI stocks, the development of AI stock indices, and the performance dynamics of AI-themed ETFs, our study fills a critical gap in the literature. Although research on these specific topics is limited, the broader themes are increasingly relevant in today's technological landscape (Claus and Krippner, 2018; Somefun et al., 2023). Our work extends the understanding of thematic ETFs, particularly those centered around emerging technologies like AI, by assessing their market performance against suitable benchmarks. Our contribution enhances the field of ETF and index performance analysis (Häusler and Xia, 2022; Schröder, 2007) and offers practical insights for investors interested in technology-driven thematic investing (Wu and Chen, 2022). Additionally, our focus on transparency and the challenges associated with thematic ETFs adds to the discourse on the risks and rewards of thematic investing (Blitz, 2021a; Raghunandan and Rajgopal, 2022), especially in sectors characterized by rapid technological evolution and market volatility. By exploring these issues, this study bridges the gap between theoretical constructs of thematic investing and the practical realities faced by investors in the dynamic AI sector. This alignment of theoretical insights with practical applications underscores the significance of our research for both academia and the investment industry.

The primary objective of this study is to develop an objective, data-driven methodology for classifying AI stocks based on publicly available corporate disclosures. By leveraging natural language processing (NLP) techniques on annual 10-K filings, we aim to quantify each company's engagement with AI technologies. Specifically, we seek to answer the following research questions: (I) How can an NLP analysis of 10-K filings be utilized to measure a company's objective engagement with AI?; (II) Do companies with higher AI engagement, as measured by our methodology, exhibit significantly different market reactions to major AI-related events, such as the launch of ChatGPT?; (III) How do AI stock indices constructed using our objective measures compare in performance to existing AI-themed ETFs and market benchmarks in terms of risk-return profiles and market responsiveness? By addressing these questions, we aim to contribute to the above-mentioned literature on thematic investing and offer practical implications for investors, asset managers, and policymakers in the rapidly evolving AI investment landscape.

This study begins in Section 2.1 by reviewing the literature on the information content and impact of public corporate disclosures, particularly 10-K filings, on market dynamics and continues in Section 2.2 by examining thematic investing and the role of ETFs in capturing investment themes like AI. In Section 3.1, we detail the data collection process for stock and exchange-traded fund returns used in our analysis; Section 3.2 describes our methodology for measuring AI engagement in stocks through natural language processing of corporate filings; and Section 3.3 outlines the construction of four AI indices based on these AI measures. Section 4.1 validates our AI stock classification by analyzing market reactions to the launch of ChatGPT; Section 4.2 assesses the predictive power of our AI indices on abnormal returns; and Section 4.3 examines abnormal returns specifically in AI-classified stocks. In Section 4.4, we compare the AI engagement in our indices with existing AI-themed ETFs, and Section 4.5 provides a comprehensive performance assessment of our AI indices and these ETFs. Section 5.1 discusses implications for theory, highlighting contributions to finance and innovation literature; Section 5.2 explores practical implications for investors, fund managers, and policymakers; and Section 5.3 addresses limitations and suggests avenues for future research. Finally, Section 6 concludes the study by summarizing our findings and emphasizing their significance for thematic investing in AI.

## 2. Literature Review

### 2.1. Information Content and Public Disclosures

The study of financial markets has extensively explored how disclosures by public companies influence market dynamics, including stock returns and trading volumes, a crucial component in understanding how markets respond to new information and how this affects investment behavior (Admati and Pfleiderer, 1988). Advancements in information technology have greatly reduced costs and enhanced the speed of data generation, aggregation, and analysis (Johnson, 2001). This shift has emphasized the importance of data processing and interpretation over mere data possession (Lee and Cho, 2005).

A notable technological advancement in this arena is the implementation of the EDGAR system, introduced in the 1990s, which has improved the timeliness and availability of corporate filings such as 10-K reports. Consequently, expertise in data management, information processing, and financial literacy has become indispensable for effective stock valuation and trading strategies (Lusardi and Mitchelli, 2007; van Rooij et al., 2011). This is further underscored by the increasing complexity and length of these filings, which often challenge investors to interpret financial information effectively (Dyer et al., 2017).

Beaver's (1968) seminal work illustrated the significant impact of earnings announcements on stock prices, revealing a market reaction to corporate financial disclosures. Later studies showed that delays or lack of transparency in filings can amplify market reactions, suggesting the importance of timely and clear disclosures in stabilizing or informing stock market behaviors (Griffin, 2003; Morse, 1981). This body of research reinforces the idea that earnings announcements and other key disclosures remain pivotal events, affecting stock price fluctuations and influencing trading activity (Beaver et al., 2020; Bonsall et al., 2020; Vamossy, 2021).

The information value of 10-K filings, particularly concerning market reactions to firms' financial disclosures, has been a significant focus of academic focus. Studies confirmed that investors deem the information content disclosed valuable (Aharony and Swary, 1980; Ball and Kothari, 1991; Bonsall et al., 2020; Noh et al., 2021; Stice, 1991), supporting the efficient market hypothesis (EMH) (Fama, 1970). These disclosures introduce new, pertinent information to investors, influencing the supply and demand dynamics that determine an asset's equilibrium price, as reflected in observable stock price fluctuations post-disclosure.

The relationship between 10-K filings and stock market responses is multifaceted. While key financial metrics like earnings per share or dividends may be anticipated or are even public knowledge before the official filing with the SEC, actual filing days are associated with abnormal stock price changes, as 10-K filings reveal additional insights into a company's future performance (You and Zhang, 2009). Research has analyzed various aspects of the information content in financial reports. For instance, this includes the impact of report readability and complexity on confidence in reported information (Rennekamp, 2012) or how these factors influence trading behavior (Miller, 2010) and bond ratings (Bonsall and Miller, 2017). This suggests that clarity in disclosures signals firm transparency and quality (Spence, 1973).

Additionally, company disclosures play a crucial role in bridging the information asymmetry between a company and its stakeholders (Campbell et al., 2014). The quality of financial reporting, therefore, becomes a pivotal factor influencing investor decisions. High-quality reporting, characterized by clarity and comprehensive detail, can attract investors by signaling robust company health. Conversely, reports that hint at potential risks or lack transparency may deter investment. In essence, the nature of financial reporting in 10-K filings not only reflects a company's current state but also shapes investor perceptions and market responses.

## 2.2. The ChatGPT Effect

Since the debut of OpenAI's ChatGPT in November 2022, which quickly became the fastest-growing consumer application on record (Hu, 2023), the debate about the capabilities of artificial intelligence (AI) technologies has gained significant traction across various sectors, particularly in financial markets. The breakthrough of large language models (LLMs) has catalyzed a global conversation regarding AI's purported potential (Bhaimiya, 2023; Murphy Kelly, 2023), resulting in a spillover effect that has elevated both the perceived potential and profitability of AI-centered investments.

The growing global focus on AI and its capabilities has significantly influenced investor behavior, as evidenced by notable stock performance in the sector. For instance, Nvidia's stock value more than quadrupled (Bowman, 2024), and AI-specialized firm c3.ai saw its stock rise by 28% following the incorporation of ChatGPT into its AI toolset (Fox, 2023). Even Buzzfeed, a digital media company not primarily associated with AI, experienced a 120% surge in its share price following the announcement of adopting OpenAI's technology for content creation (Diaz and Smith, 2023). These examples reflect a broader market trend where companies benefit from their association with AI, regardless of their direct involvement with ChatGPT or LLMs. Further illustrating this trend, companies like BigBear.ai and SoundHound AI saw significant stock gains following the launch of ChatGPT (Singh and Biwas, 2023), highlighting increased investor confidence in AI-focused businesses. However, this surge in AI-related investments has also led to regulatory scrutiny. Additionally, research on crypto assets has shown that AI-centric assets have experienced significant abnormal positive price movements due to the heightened perception and valuation of AI's potential post-ChatGPT (Ante and Demir, 2024; Saggu and Ante, 2023). This spillover effect underscores the profound impact of AI advancements on stock market dynamics and investor sentiment.

Investor enthusiasm for AI predates the era of ChatGPT. Research has shown that exchange-traded funds (ETFs) with "AI" in their names were valued approximately 0.4% higher than comparable ETFs without the term AI, highlighting an AI name premium (Wu and Chen, 2022). Recent advancements in LLMs have extended this early recognition of AI's (perceived) economic significance. It is important to acknowledge that LLMs and the broader field of generative AI—which, e.g., includes the creation of pictures, music, or videos—represent just one application area of AI that has captured significant public interest. Nevertheless, the success of generative AI does not detract from AI's potential in other domains. In fact, the achievements in generative AI likely contribute to a spillover effect, enhancing the perceived overall potential of AI across different sectors.

Given the rising influence of AI in financial markets, corporate disclosures of AI will likely become critical in shaping investor perceptions and decision-making. The substantial stock performance boosts observed following AI-related announcements underscore the market's demand for transparent and reliable AI reporting—also to prevent "AI washing" (SEC, 2024). As companies increasingly integrate

AI, the frequency and clarity of their AI disclosures could serve as valuable indicators of technological engagement, offering investors a clearer view of a company's long-term growth potential and innovative capabilities. The market's response to AI mentions—as evidenced by significant stock price reactions—would suggest that consistent, well-defined AI disclosures enhance investor confidence and allow for more precise valuation assessments. Thus, as AI continues to redefine industries, accurate reporting of AI involvement in corporate filings will likely become a cornerstone of informed investment strategies, providing investors with essential insights into a company's technological trajectory and competitive positioning in an AI-driven world.

## 2.3.   Technology and AI in Public Disclosures

Corporate filings, particularly 10-K reports, have long been a key source for extracting firm-level insights, with recent analyses focusing on technology-related disclosures due to their potential impact on stock performance. For example, Andreou et al. (2020) employed textual analysis to formulate a measure of market orientation and uncover a positive correlation with firm performance. Research has identified that technology-specific announcements, including those on blockchain and the metaverse, often prompt positive market reactions, underscoring investor receptiveness to technological advancements as indicators of future growth potential(Aharon et al., 2022; Cahill et al., 2020; Cheng et al., 2019).

In parallel, the recent literature has focused on the analysis of sustainability-related topics, notably corporate social responsibility (CSR) (Cannon et al., 2020) and environmental, social, and governance (ESG) issues (Jiang et al., 2023). For instance, Baier et al. (2020) highlighted that an average of 4% of all words in 10-K reports pertained to ESG, signifying the growing importance of these issues in corporate disclosure and communication, a sentiment echoed by Heichl and Hirsch (2023). The tone used in communicating ESG matters within 10-K filings has been linked to stock market performance around the filing dates (Ignatov, 2023). This underscores the significance not just of the content but also of the manner in which it is conveyed, reinforcing the idea that the qualitative aspects of corporate communications are pivotal in shaping market reactions.

Amid these disclosure trends, AI-related information has emerged as a new focal point, yet research in this area remains limited. Mishra et al. (2022) found that AI usage in firms, as identified through textual analysis, correlates with higher profitability and efficiency. On the other hand, Wang and Yen (2023) discovered that firms disclosing AI involvement tend to have higher stock prices, and AI-related risk disclosures are viewed positively by investors. While both studies have clear relevance to this study, they pursue significantly different objectives.

## 2.4. Thematic Investing and ETFs

As investors increasingly seek exposure to companies actively engaged in transformative technologies like AI or in adhering to ESG standards, thematic investing has gained popularity as a strategy to target long-term trends that influence the global economy. This investment approach centers on capitalizing on emerging areas such as technological innovation, demographic shifts, and environmental changes (Claus and Krippner, 2018), thus aligning with the growing emphasis on thematic corporate disclosures explored above. The growing popularity of thematic investment strategies reflects investors' growing interest in high-growth areas and innovative industries, as thematic investing aligns portfolios with these dynamic shifts (Somefun et al., 2023). sectors positioned to benefit from these overarching trends, thematic investing represents a forward-looking approach that increasingly relies on corporate disclosures of sector-specific engagement, like technology adoption or sustainability practices (Dheeriya and Malladi, 2019; Methling and von Nitzsch, 2020, 2019).

Thematic ETFs, including those focused on technology, AI, or ESG, serve as prime examples for thematic investing by giving investors direct access to specific market segments without the need to handpick individual stocks. Technology ETFs, for example, include investments ranging from established technology companies to emerging fields like AI, robotics, and biotech (Wu and Chen, 2022). These ETFs align with thematic investment goals by focusing on sectors expected to drive economic growth and innovation, providing investors with a diversified entry point into the companies leading these trends. Thematic ETFs promise to simplify access to specific market segments by offering targeted exposure without the need for investors to handpick individual stocks. This feature is especially appealing to investors who want exposure to trends but lack the resources or expertise to identify or analyze individual companies engaged in these areas.

However, thematic investing presents unique challenges related to transparency in asset selection. A primary concern is the opacity in asset selection criteria for these funds, which raises questions about how closely their portfolios truly align with advertised themes. This lack of clarity complicates the evaluation process for investors, who may find it difficult to determine whether a thematic ETF genuinely reflects the underlying trends it claims to target (Raghunandan and Rajgopal, 2022). For instance, while technology-themed ETFs promote exposure to innovation, the predictive nature of thematic investing inherently carries speculative risks due to its reliance on forward-looking assumptions about technological trends and adoption rates (Blitz, 2021a). Ben-David et al. (2023) highlight that specialized ETFs tend to underperform over their first five years, with about a 30% risk-adjusted loss. This underperformance is linked to the overvaluation of underlying stocks at the time of the launch, suggesting that providers cater to investors' extrapolative beliefs by issuing ETFs that focus on attention-grabbing themes rather than delivering sustainable value. These factors collectively underscore the importance of investor diligence and awareness of the speculative and volatile nature inherent in thematic investing within the dynamic technology sector.

For the topic of AI, Bonaparte (2023) introduced an exemplary valuation model for AI stocks and ETFs, utilizing the correlation between AI-related Google Trends sentiment and stock performance. This innovative approach underscores the potential for alternative, sentiment-driven measures in evaluating AI-focused ETFs, yet differs significantly from the methodology in this study. Here, we address the opacity and misalignment challenges in thematic ETFs by introducing a data-driven framework based on corporate disclosures designed to identify and quantify AI engagement among firms objectively. Our approach aims to provide investors with a more transparent and reliable method for assessing AI-themed assets, ensuring that these ETFs align closely with genuine technological involvement rather than speculative trends. By focusing on objective metrics derived from firm-level disclosures, this study aims to make thematic investing in technology sectors, especially AI, more efficient and relevant to long-term investor goals.

## 3.  Data Background and Methodology

### 3.1.  Stock and Exchange-Traded Fund Returns

To construct our empirical dataset, we begin by extracting a complete list of all 4,316 assets listed on the Nasdaq stock exchange from January 1, 2010, to September 1, 2022, using the Nasdaq stock screener tool. [1] [2] The Nasdaq was specifically selected due to its historical preference for a more accommodating listing of technology firms, a factor that contributes to its reputation as a primarily technology-oriented exchange. This is evident in its substantial weighting towards technology companies, including those specializing in artificial intelligence (Nasdaq, 2024). For each stock, we obtained the historical daily closing price and Central Index Key (CIK) from the S&P Capital IQ database.[3] Following a thorough data-cleaning process removing non-stock assets, we obtained a sample of 3,470 stocks. After mapping each asset to its CIK, we derive a final sample of 3,395 stocks.

---

[1] Different online data sources, including the Financial Times, Wall Street Journal, Bloomberg, and Thomson Reuters Eikon, report varying numbers of assets listed on the Nasdaq. To address these inconsistencies and ensure accuracy, we consulted Nasdaq directly. The Nasdaq recommended using their Stock Screener tool as the definitive source of data. This tool, which enables the filtering and analysis of stocks, is accessible online at https://www.nasdaq.com/market-activity/stocks/screener.

[2] The complete list of assets comprises those from the Nasdaq Global Select Market, which includes large-cap companies meeting the most stringent financial and liquidity standards; the Nasdaq Global Market, featuring mid-cap companies that satisfy moderately rigorous requirements; and the Nasdaq Capital Market, which lists small-cap companies meeting baseline entry criteria. This stratification ensures a diverse representation of companies across different market capitalizations and financial thresholds.

In addition to sourcing Nasdaq stocks, we utilized the VettaFi ETF database to identify AI-related ETFs and conducted a targeted online search using key phrases such as "AI ETF," "AI stocks," and "invest in artificial intelligence" to capture a broader range of relevant AI-focused funds. Our resulting sample comprises 14 of the largest AI ETFs available. The list is not intended to be exhaustive but rather representative of the largest AI ETFs available. Table 1 documents each ETF's ticker symbol, full name, expense ratio, and asset count.[4] The last column includes an excerpt of the high-level logic from each fund's documentation, summarizing the types of assets included based on a review of each ETF's website.

A preliminary analysis reveals several issues. First, many AI-focused ETFs include related themes such as robotics, big data, and deep learning rather than solely focusing on AI. Secondly, the criteria for asset inclusion are often vague and nonspecific. For instance, many ETF providers use broad language in their descriptions of asset selection criteria, such as including companies "that potentially stand to benefit from increased adoption and utilization of artificial intelligence." In many cases, the ETFs are using an arbitrary selection of stocks that they consider AI-related rather than relying on any real objective metric. This finding reinforces the critical role of our research in establishing an objective metric to address the arbitrariness observed in current ETF asset selection.

For each ETF, we collect the historical daily closing prices and CIK, along with those of the Nasdaq Composite Index (IXIC) and the S&P 500 Index (SPX), for benchmarking purposes, from S&P Capital IQ. The daily prices for 3,395 stocks, 14 AI-related ETFs, SPX, and IXIC are subsequently converted into daily logged returns.

---

[3] The Central Index Key (CIK) is a unique identifier assigned by the U.S. Securities and Exchange Commission (SEC) to entities that file disclosures and reports with the SEC, such as public companies and investment funds. The CIK distinguishes between entities and facilitates the retrieval of public filings from the SEC's EDGAR (Electronic Data Gathering, Analysis, and Retrieval) database. It ensures that each entity's filings can be accurately tracked and accessed.

[4] Figure A.1 illustrates a scatter plot comparing the expense ratios and mean daily returns of the AI-themed ETFs listed in Table 1. Each data point represents an individual ETF, plotted according to its expense ratio and corresponding mean daily return over the specified period. This visual representation aids in analyzing the cost-efficiency of these ETFs, providing empirical insight into the relationship between the costs incurred by investors and the returns generated. Notably, the AI-themed ETFs exhibit expense ratios significantly higher than 0.25%, positioning them at the higher end of the ETF pricing spectrum. Our analysis does not reveal a positive correlation between expense ratios and returns during the period under consideration, suggesting that higher fees do not necessarily translate into better performance in the AI-themed ETF market.

**Table 1. Overview of artificial intelligence-related exchange-traded funds**

| Ticker | Full name | Inception | Expense ratio | No. of assets | Asset selection criteria excerpt |
|---|---|---|---|---|---|
| ROBO | Robo Global Robotics and Automation Index ETF | 09/12/16 | 0.95 | 78 | "…companies that are driving transformative innovations in robotics, automation, and artificial intelligence…" (ROBO GLOBAL, 2023a) |
| BOTZ | Global X Robotics & Artificial Intelligence ETF | 10/21/13 | 0.69 | 44 | "…companies that potentially stand to benefit from increased adoption and utilization of robotics and artificial intelligence…" (Global X, 2023a) |
| ROBT | First Trust Nasdaq Artificial Intelligence and Robotics ETF | 02/21/18 | 0.65 | 109 | "…companies engaged in Artificial intelligence (AI), robotics and automation […] as determined by the Consumer Technology Association (CTA)…" (First Trust, 2023) |
| ARKQ | ARK Autonomous Technology & Robotics ETF | 09/30/14 | 0.75 | 36 | "at least 80% of its assets […] securities of autonomous technology and robotics companies…" (ARK Invest, 2023) |
| AIQ | Global X Artificial Intelligence & Technology ETF | 05/11/18 | 0.68 | 87 | "…companies that potentially stand to benefit from the further development and utilization of artificial intelligence (AI)…" (Global X, 2023b) |
| IRBO | iShares Robotics and Artificial Intelligence Multisector ETF | 05/26/18 | 0.47 | 112 | "…companies that could benefit from the long-term growth and innovation in robotics technologies and artificial intelligence…" (iShares, 2023) |
| IGPT | Invesco AI and Next Gen Software ETF | 06/22/05 | 0.61 | 100 | "…at least 90% of its total assets […] companies with significant exposure to technologies or products that contribute to future software development through direct revenue…" (Invesco, 2023) |
| XAIX | Xtrackers Artificial Intelligence & Big Data UCITS ETF | 01/29/19 | 0.35 | 94 | "…companies […] that have material exposure to themes related to among other AI, big data and cyber security that meet certain ESG Criteria." (DWS, 2023) |
| GOAI | Amundi MSCI Robotics & AI ESG Screened UCITS ETF | 09/04/19 | 0.40 | 157 | "…companies associated with the increased adoption and utilization of artificial intelligence, robots and automation." (MSCI, 2023) |
| WTAI | WisdomTree Artificial Intelligence and Innovation Fund | 12/09/21 | 0.45 | 77 | "…companies that are primarily involved in the investment theme of AI and Innovation." (WisdomTree, 2023) |
| QTUM | Defiance Quantum ETF | 09/04/18 | 0.40 | 69 | "…companies that derive at least 50% of their annual revenue or operating activity from the development of quantum computing and machine learning technology" |
| AIAG | L&G Artificial Intelligence UCITS ETF | 07/02/19 | 0.49 | 62 | "…companies developing the technology and infrastructure enabling AI…" (ROBO GLOBAL, 2023b) |

## 3.2. Measuring AI in Stocks

To measure the extent to which each of the 3,395 stocks is related to AI, we firstly begin by utilizing the advanced search function of the SEC's EDGAR (Electronic Data Gathering, Analysis, and Retrieval) database to collect the annual 10-K filings for each stock over the sample period. The 10-K filings are comprehensive reports that publicly traded companies file annually to disclose their financial performance. Mandated by the SEC, these filings provide detailed information on a company's financial health, including earnings, expenses, assets, liabilities, and other relevant metrics. They serve as essential tools for investors and analysts, offering deep insights into a company's operations, market position, and strategic direction (Mishra et al., 2022). Notably, the 10-K filings often contain references to the company's engagement with emerging technologies like AI, shedding light on their strategic priorities and innovation initiatives.

Secondly, the text of each filing was cleaned and tokenized. This process, which is a fundamental preprocessing step in natural language processing (NLP), involves breaking down the text into smaller units called tokens, such as words or phrases. This helps in analyzing the text more effectively. For instance, the phrase 'we utilize artificial intelligence' would be tokenized into individual words: ['we', 'utilize', 'artificial', 'intelligence']. This step ensures that our analysis is based on a clear and manageable set of data.

Thirdly, to systematically quantify AI-related activity for each stock, specifically in each company's 10-K filings, we conducted a bibliometric analysis using the Web of Science (WoS) database to identify pertinent AI-related terms. Specifically, we used the search string $TS="artificial intelligence*"$ to extract the 500 most recent WoS articles containing this term in their titles or abstracts. Using VOSviewer software (van Eck and Waltman, 2010), we derived the most common keywords, which included general terms such as *"classification" (N = 20)* and *"prediction" (N = 10)* but also more related keywords such as *"machine learning" (N = 53), "deep learning" (N = 27), "neural network*" (N = 21)* or *"blockchain" (N = 6)*. However, to maintain focus on AI-specific disclosures, we restricted our analysis to the following core keywords: (1) *"artificial intelligence*,"* (2) *"ai,"* and (3) *"a.i."*. These keywords formed the foundation of our analysis, enabling us to construct an objective measure of AI involvement across firms based on their 10-K public disclosures.

Fourthly, we calculated the frequency of each AI-related keyword within the SEC filings $C_i$ to quantitatively measure the degree of AI engagement for each firm (Equation 1):

$$C_i = \sum_{i=1}^{N} freq(k_i), \qquad\qquad\qquad (1)$$

where $freq(k_i)$ represents the frequency of the $k$-th keyword in the $i$-th filing, and $N$ is the number of keywords. By quantifying the frequency of AI-related terms, we capture the level of emphasis each company places on AI in their public disclosures. This approach provides an objective way to measure AI engagement, allowing us to systematically compare firms. In addition to the frequency measure, we generated a binary AI dummy variable that indicates whether AI-related terms are present in a given year's filing. This binary classification helps to easily identify which firms are explicitly mentioning AI, providing a clearer and more detailed picture of how widespread AI adoption is across the sample.

Figure 1 shows a gradual increase in AI mentions in 10-K filings by NASDAQ-listed firms from 2010 to 2015, with filings ranging from 7 to 11 per year. From 2016 onwards, the mentions began to rise from 37 in 2016 to 86 in 2017 and continued upwards with notable increases in 2018 (134 filings), 2019 (187 filings), and 2020 (239 filings). The most substantial growth occurred in 2021 and 2022, reaching 442 and 527 filings, respectively. This trend highlights the growing significance of AI in corporate strategy and risk disclosures, indicating broader adoption and integration of AI technologies by NASDAQ-listed firms in recent years.

**Figure 1. Number of 10-K filings of NASDAQ firms mentioning AI per year**

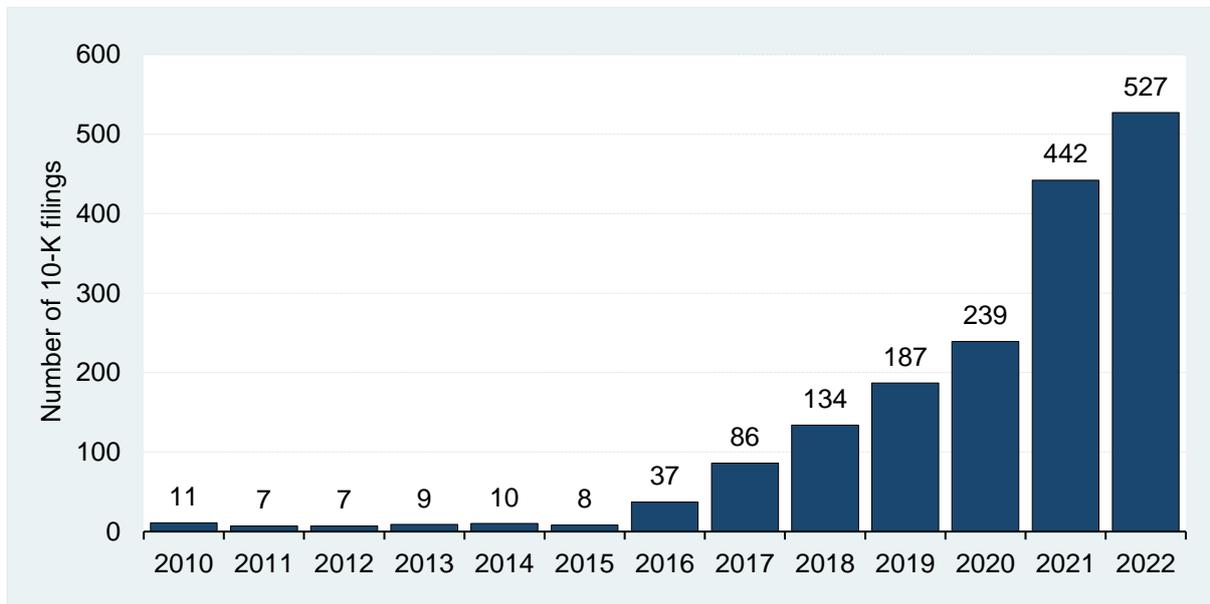

Note: Figure 1 shows the number of NASDAQ-listed companies with annual 10-K filings that mention artificial intelligence (AI) from 2010 to 2022, based on a dataset of 3,395 assets related to stocks, excluding non-stock assets.

Fifth, to further refine our metrics, we adjusted for document length to prevent longer 10-K filings from biasing the results. Specifically, we identified the most common word ($w_i$) within each filing ($M_i$):

$$M_i = \max(freq(w_i)). \tag{2}$$

For a more nuanced measure, we calculated the Term Frequency-Inverse Document Frequency (TF-IDF). TF-IDF is a commonly used metric in the analysis of large texts that gauge the importance of a word to a document in a collection or corpus (Beel et al., 2016). The TF-IDF value increases proportionally to the number of times a word appears in the document but is offset by the frequency of the word in the corpus, which helps to adjust for the fact that some words appear more frequently in general (Sparck Jones, 1972). The TF component measures how frequently a term appears in a document. In contrast, the IDF component adjusts for the term's frequency across all documents, thus helping to highlight particularly informative terms.

The term frequency ($TF_{t,d}$) for a term $t$ in a document $d$ can be defined as:

$$TF_{t,d} = \frac{f_{t,d}}{\sum_{t' \in d} f_{t',d}}. \tag{3}$$

The inverse document frequency $IDF_{t,\Delta}$ for a term $t$ in a corpus $\Delta$ is computed as:

$$IDF_{t,\Delta} = \log\left(\frac{|\Delta|}{|\{d \in \Delta : \in d\}|}\right). \tag{4}$$

where $|\Delta|$ is the total number of documents in the corpus (i.e., the number of 10-K filings in a period under consideration – in the case of annual filings = 1), and $|\{d \in \Delta : \in d\}|$ is the number of documents where the term $t$ appears. This adjustment helps to reduce the weight of terms that are common across many documents and increase the weight of more unique terms. The TFIDF score, $TF\text{-}IDF_{t,d}$, for term $t$ within document $d$ is thereby calculated as:

$$TF\text{-}IDF_{t,d} = TF_{t,d} \times IDF_{t,\Delta}, \tag{5}$$

Finally, we normalize the TF-IDR score by dividing it by the total number of words in a document ($L_d$), yielding a weighted AI score per stock per period:

$$TF\text{-}IDF_{norm,t,d} = \frac{TF\text{-}IDF_{t,d}}{L_d}. \tag{6}$$

This normalized TF-IDF score allows us to objectively compare the prominence of AI-related terms across different firms, regardless of document length. In summary, we derived both a binary AI dummy variable per stock and a weighted AI score per stock, providing a comprehensive view of AI engagement. The binary variable indicates whether a company mentions AI, while the weighted score reflects the extent and significance of those mentions. This dual approach enables a deeper understanding of the degree to which firms are engaged with AI technology, allowing for more informed comparisons and analyses across the sample.

## 3.3. Construction of AI Indices

We construct four distinct AI stock indices, including the Equally Weighted AI Index (AII), Size-Weighted AI Index (SAII), and two Time-Discounted AI Indices (TAII05 and TAII5X), which reflect the intensity and persistence of AI-related disclosures in corporate annual filings, utilizing the metrics developed in section 3.3. These indices are developed to provide investors with a more objective and data-driven approach to evaluating corporate AI engagement and its impact on market performance, ensuring an unbiased and reliable evaluation.[5]

The **Equally Weighted Artificial Intelligence Stock Index (AII)** serves as a baseline measure, incorporating all companies that mention AI in their latest (i.e., last year's) annual filings, regardless of the extent of their communication.[6] This approach assumes equal potential impact from any AI disclosure, providing a broad view of the market's engagement with AI without giving undue emphasis to any individual firm's disclosure intensity. The index provides a simple aggregation of the overall market's AI engagement.

The weight assigned to each company $i$ in year $t$, denoted as $w_{i,t}$ is defined as $\frac{1}{N_t}$ if the company mentions AI in their latest annual filing, where $N_t$ is the total number of companies. The AII is thereby formulated as:

$$AII_t = \sum_{i=1}^{N_t} \frac{1}{N_t} \times V_{i,t} \,, \tag{7}$$

where $V_{i,t}$ is the market capitalization of company $i$ in year $t$.

The **Size-Weighted Artificial Intelligence Stock Index (SAII)** differentiates companies based on the extent of their AI communication. The underlying rationale is that companies providing more detailed and extensive AI disclosures are more likely to invest more resources into AI, demonstrating a deeper engagement and a stronger commitment to AI initiatives. This deeper involvement justifies assigning these companies a higher weight in the index, reflecting a more meaningful role in AI.

The weight assigned to each company $i$ in year $t$, is defined as $\frac{S_{i,t}}{\sum_{j=1}^{N_t} S_{i,t}}$, where $S_{i,t}$ denotes the AI score of company $i$ in year $t$. The SAII is thereby formulated as:

$$SAII_t = \sum_{i=1}^{N_t} \left( \frac{S_{i,t}}{\sum_{j=1}^{N_t} S_{i,t}} \right) \times V_{i,t} \,. \tag{8}$$

The **Time-Discounted Artificial Intelligence Stock Index (TAII)** introduces a temporal dimension, capturing both current and historical AI communications to provide a more comprehensive understanding of corporate AI engagement over time. By incorporating the historical context, the TAII aims to assess both the immediate impact of AI disclosures and how sustained or past AI engagement influences market perception and performance. We construct two variants of TAII: one with a discount factor ($\alpha$) of 0.5, reflecting a moderate decay in the relevance of past communications (TAII05), and another with $\alpha$ set to 5, reflecting an increasing impact of historical AI communication (TAII5X). These variants allow us to examine how different levels of emphasis on past AI activities can potentially influence market analysis and the overall index performance.

The TAII for a given $\alpha$ is therefore calculated as:

$$TAII_t = \sum_{i=1}^{N_t} \left( \frac{D_{i,t}}{\sum_{j=1}^{N_t} D_{i,t}} \right) \times V_{i,t} \,, \tag{9}$$

where $D_{i,t}$ represents the discounted sum of past AI engagements for company $i$, defined as $D_{i,t} = \Theta_{i,t} + \alpha \times \Theta_{i,t-1} + \alpha^2 \times \Theta_{i,t-2} + \cdots + \alpha^n \times \Theta_{i,t-n}$ and $\Theta_{i,t}$ is a dummy variable indicating if a company $i$ had an AI score in year $t$. The discount factor $\alpha$ determines how much weight is given to historical AI communications. Higher values of $\alpha$ emphasize the importance of past disclosures, while a lower discount factor ($\alpha$) indicates that older AI communications are less relevant.

All indices are updated daily to reflect market dynamics, ensuring that the indices remain responsive to ongoing changes in AI engagement and stock performance. The indices are also rebalanced annually to maintain consistency and alignment with the latest available data. The value of each index on day $t$, denoted as $I_t$, is recalculated based on the previous day's index value $I_{t-1}$ and the daily returns $R_{i,t}$ of the constituent stocks:

$$I_t = I_{t-1} \times \left(1 + \sum_{i=1}^{N_t} w_{i,t} \times R_{i,t}\right). \tag{10}$$

Our methodology provides a clear, multifaceted, data-driven approach to defining and constructing AI indices, in contrast to the arbitrary approaches often employed by ETFs to create AI-themed funds. The AII offers a broad market perspective, focusing on the presence of AI disclosures. The SAII provides a deeper analysis by weighing companies based on the extent of their AI engagement, thereby highlighting the intensity of corporate commitment to AI initiatives. The TAII, with its temporal emphasis, captures the evolving narrative of AI involvement, enabling a more dynamic assessment of how past and present AI communications influence investor perception. Together, these indices offer a comprehensive toolset for analyzing corporate AI communication, providing investors with an objective framework for understanding AI engagement in the market.

Figure 2 reveals that all AI indices (AII, SAII, TAII05, and TAII5X) consistently outperformed the NASDAQ Composite Index (IXIC) from 2011 to 2023, highlighting the value of AI-focused investment strategies. Among these indices, TAII5X demonstrated the highest cumulative return, highlighting the benefits of heavily weighting historical AI disclosures when assessing long-term corporate AI engagement. While TAII05 and SAII also showed strong performance, they were slightly below TAII5X. The AII, while still surpassing the IXIC, showed comparatively more moderate growth, suggesting that assigning equal importance to all AI mentions captures broad engagement but may overlook the impact of deeper, sustained AI involvement. This comparison highlights the potential for better market performance with sustained or historical AI involvement, encouraging readers to consider this approach in their investment strategies.

**Figure 2. Performance of AI-related stock indices versus the benchmark Nasdaq Composite Index**

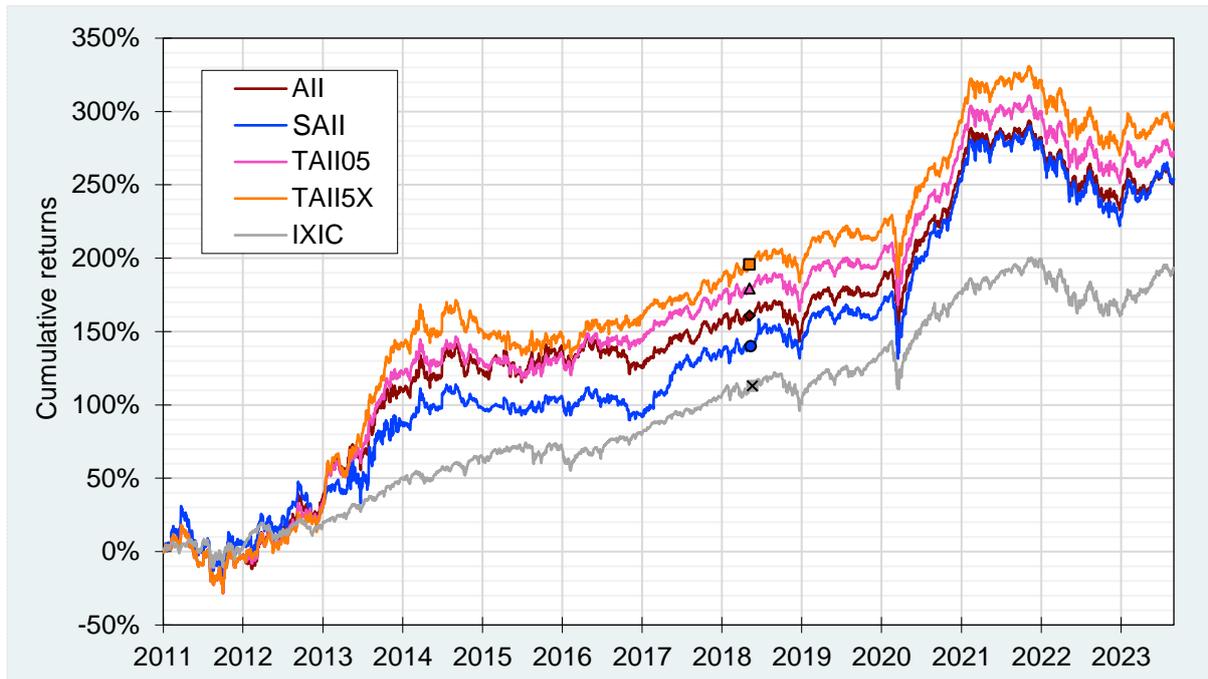

Note: Figure 2 illustrates the cumulative returns of four AI stock indices—Equally Weighted AI Index (AII), Size-Weighted AI Index (SAII), Time-Discounted AI Index with a discount factor of 0.5 (TAII05), and Time-Discounted AI Index with a discount factor of 5 (TAII5X)—in comparison to the benchmark NASDAQ Composite Index (IXIC) from January 3, 2011, to September 1, 2023.

## 4. Empirical Model and Results

### 4.1. Market Reaction as a Validation of AI Stock Classification

Mentioning AI in a company's 10-K filing does not necessarily indicate active AI utilization or qualify the company as an AI-focused stock. To assess the accuracy of our AI stock classification measures in identifying a company's engagement with AI, we initiate the empirical investigation by explicitly testing their response to the launch of ChatGPT. Previous research in crypto asset markets (Ante and Demir, 2024; Saggu and Ante, 2023) demonstrates that the launch of ChatGPT on November 30, 2022, generated significant interest in AI, resulting in greater cumulative average abnormal returns (CAARs) for AI-focused crypto assets compared to non-AI crypto assets. We hypothesize that a similar effect occurred in stock markets, leading to positive abnormal returns for AI-focused stocks. While asset managers who select AI-themed ETF compositions may not necessarily possess deep technological expertise and may use arbitrary inclusion criteria, individual investors with specialized knowledge of specific companies are more likely to accurately assess a company's AI involvement, which should be reflected in their response to the ChatGPT event.

To evaluate this hypothesis, we analyze the returns of 2,541 stocks—classified as either AI or non-AI—to test (H$_0$) if AI stocks exhibit similar CAARs to non-AI stocks after the ChatGPT launch, then our NLP-based AI indices do not provide additional value. Conversely, (H$_1$) significantly different CAARs between AI and non-AI stocks indicates varying levels of AI engagement, thereby reinforcing the effectiveness of our data-driven AI indices. For this event study, we employ a market model to estimate expected returns, accounting for each stock's individual risk characteristics:

$$R_{i,t} = \alpha_i + \beta_i R_{m,t} + \varepsilon_{i,t}, \tag{11}$$

Where $R_{i,t}$ is the return of stock or AI index $i$ on day $t$, $R_{m,t}$ represents the return of the reference market—the S&P 500 Index (SPX)—on day $t$, $\alpha_i$ is the stock-specific intercept, $\beta_i$ measures the sensitivity of the stock or AI index return to market movements, and $\varepsilon_{i,t}$ is the error term. We utilize an estimation window of one year, equivalent to the 251 trading days, before the ChatGPT launch to estimate model parameters and calculate expected returns, consistent with event studies (Campbell et al., 2010). The event window spans three months, equivalent to 61 trading days, following the ChatGPT launch on November 20, 2022, consistent with the ChatGPT effect identified in previous studies (Ante and Demir, 2024; Saggu and Ante, 2023). This period allows us to capture both immediate and lagged market reactions to the event. To assess the statistical significance of the results, we employ both parametric t-tests and nonparametric Wilcoxon signed-rank tests (Wilcoxon, 1945).

Figure 3 presents the CAARs for all 2,541 Nasdaq stocks over the event window, divided into two panels. Panel (a) consists of 582 stocks classified as 'AI stocks' in 2023 based on the TAII05 index, while panel (b) comprises the remaining 1,959 stocks not classified as 'AI stocks.' The CAARs are positive and statistically significant for both groups one month following the ChatGPT launch: 17.25% for AI stocks (t = 8.24; z = 9.79) and 11.59% for non-AI stocks (t = 10.55; z = 10.83). The 5.68% difference is statistically significant at the 5% level based on a two-sample t-test, indicating that AI stocks had a stronger positive reaction to the ChatGPT launch compared to non-AI stocks. By demonstrating a statistically significant difference in the market reactions between AI and non-AI stocks, the results bolster the validity of our AI classification methodology. The stronger positive abnormal returns for AI stocks suggest that investors recognize and respond to the AI engagement of companies as captured by our NLP-based measures. This outcome aligns with our hypothesis that the launch of ChatGPT would have a more pronounced effect on stocks genuinely involved in AI activities, thus providing empirical support for the effectiveness of our data-driven AI indices.

**Figure 3. CAARs for AI and non-AI Nasdaq stocks following ChatGPT launch**

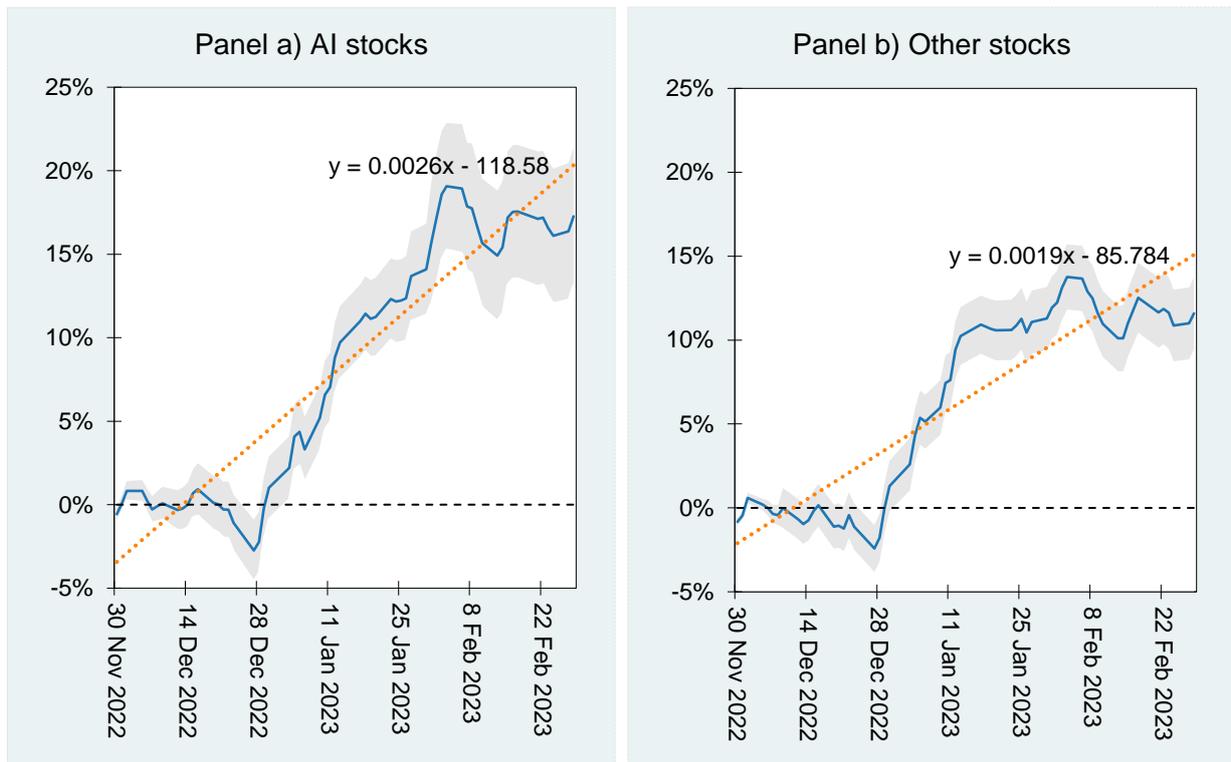

Note: Figure 3 displays the cumulative average abnormal returns (CAARs) over a 61-trading-day (three-month) event window following the launch of OpenAI's ChatGPT on November 30, 2022. Panel (a) illustrates the CAARs for 582 stocks classified as AI-related based on the TAII05 index, while Panel (b) depicts the CAARs for the remaining 1,959 non-AI stocks. The estimation window comprises 251 trading days (one year) prior to the event. The market model employs the S&P 500 Index (SPX) as the reference market. The gray bands represent the 95% confidence intervals, and the orange dotted line indicates the linear trend.

## 4.2. The Predictive Power of AI Indices on Abnormal Returns

Building upon the statistically significant differences observed between AI and non-AI stocks in Figure 3, we extend our analysis to examine how effectively our AI indices explain the CAARs following the launch of ChatGPT. Table 3 presents the regression results predicting CAARs for all 2,541 NASDAQ-listed stocks in our sample. The primary objective is to evaluate whether the AI indices described in Section 3.3 serve as significant predictors of the abnormal returns observed after this event.

$$CAAR_i = \alpha_i + \beta_i AI\_Index\_Weight_i + \varepsilon_i, \qquad (12)$$

Where $CAAR_i$ represents the cumulative average abnormal return for stock $i$, $AI\_Index\_Weight_{i,t}$ denotes the weight of different AI indices—such as AII, TAII05, and TAII5X—for the years 2022 or 2023, and $\varepsilon_i$ is the error term. Models with odd numbers (1, 3, 5, 7, 9, and 11) employ ordinary least squares (OLS) estimation, while even-numbered models (2, 4, 6, 8, 10, and 12) use robust MM estimation to account for potential outliers and heteroskedasticity. Models 1 and 2 use the Equally Weighted Artificial Intelligence Stock Index (AII) for 2022. Models 3 and 4 use the AII for 2023. Models 5 and 6 employ the Time-Discounted Artificial Intelligence Stock Index with a discount factor of 0.5 (TAII05) for 2022. Models 7 and 8 use the TAII05 for 2023. Models 9 and 10 utilize the Time-Discounted Artificial Intelligence Stock Index with a discount factor of 5 (TAII5X) for 2022. Models 9 and 10 utilize the Time-Discounted Artificial Intelligence Stock Index with a discount factor of 5 (TAII5X) for 2022.

The estimates in Table 2 reveal that the coefficients for the AI index weights are positive and statistically significant across all models. This finding reinforces the reliability of our AI indices in predicting the abnormal returns observed after the ChatGPT launch. Specifically, the coefficients are largest for the AII metric, followed by the TAII05 metrics, and smallest for the TAII5X metrics. This gradient suggests that indices capturing more recent AI engagement (AII and TAII05) have a stronger association with abnormal returns compared to those placing greater emphasis on historical AI disclosures (TAII5X) in 10-K filings. Additionally, the coefficients for the 2022 index weights are consistently higher than those for the 2023 weights across corresponding models. This pattern indicates that investors placed greater emphasis on prior AI-related disclosures when reacting to the ChatGPT launch, highlighting the importance of recent AI communications in shaping market responses. Overall, the results demonstrate that our AI indices effectively capture the differential impacts of each company's AI engagement—as revealed through 10-K disclosure—on stock performance. Stocks with higher AI index weights— indicative of greater AI involvement—exhibited stronger abnormal returns in response to the ChatGPT event. Our estimates reinforce the validity of our NLP-based AI stock classification methodology and helps explain market behavior in the context of significant AI developments.[7]

---

[7] In line with MacKinlay (1997), low R-squared values are not necessarily a limitation in event studies as the analysis focuses on the significance of the abnormal return rather than the overall explanatory power. The low R-squared values indicate that numerous factors beyond our AI indices influence stock returns, reflecting the complexity of market dynamics. However, the statistical significance of the coefficients suggests that our AI indices do capture meaningful variations in cumulative average abnormal returns associated with AI engagement. Despite the modest explanatory power, the significant coefficients provide evidence that our AI indices are relevant predictors of stock performance in response to the ChatGPT launch, thereby validating the utility of our NLP-based classification methodology.

**Table 2. Regression predicting CAARs for all Nasdaq stocks after ChatGPT launch**

| | (1) | (2) | (3) | (4) | (5) | (6) | (7) | (8) | (9) | (10) | (11) | (12) |
|---|---|---|---|---|---|---|---|---|---|---|---|---|
| AII (2022) | 0.068** | 0.053*** | | | | | | | | | | |
| | (0.027) | (0.015) | | | | | | | | | | |
| AII (2023) | | | 0.052** | 0.050*** | | | | | | | | |
| | | | (0.026) | (0.015) | | | | | | | | |
| TAII05 (2022) | | | | | 0.050*** | 0.040*** | | | | | | |
| | | | | | (0.018) | (0.010) | | | | | | |
| TAII05 (2023) | | | | | | | 0.040** | 0.035*** | | | | |
| | | | | | | | (0.016) | (0.009) | | | | |
| TAII5X (2022) | | | | | | | | | 0.015*** | 0.012*** | | |
| | | | | | | | | | (0.006) | (0.003) | | |
| TAII5X (2023) | | | | | | | | | | | 0.012*** | 0.009*** |
| | | | | | | | | | | | (0.005) | (0.003) |
| Constant | 0.118*** | 0.051*** | 0.120*** | 0.050*** | 0.117*** | 0.049*** | 0.118*** | 0.049*** | 0.120*** | 0.051*** | 0.118*** | 0.050*** |
| | (0.011) | (0.006) | (0.011) | (0.006) | (0.011) | (0.006) | (0.011) | (0.009) | (0.010) | (0.006) | (0.011) | (0.006) |
| R2 / Pseudo R2 | 0.002 | 0.003 | 0.002 | 0.003 | 0.010 | 0.004 | 0.002 | 0.004 | 0.003 | 0.004 | 0.003 | 0.003 |
| F / Chi2 | 0.0106 | 0.0005 | 0.0403 | 0.0007 | 0.0070 | 0.0000 | 0.0132 | 0.0001 | 0.0096 | 0.0001 | 0.0104 | 0.0003 |
| Model | OLS | MM | OLS | MM | OLS | MM | OLS | MM | OLS | MM | OLS | MM |

Note: Table 2 reports estimates from regressions predicting cumulative average abnormal returns (CAARs) for all 2,541 NASDAQ-listed stocks following the ChatGPT launch on November 30, 2022, as specified in Equation (12). Models 1, 3, 5, 7, 9, and 11 present ordinary least squares (OLS) estimates, while Models 2, 4, 6, 8, 10, and 12 provide robust MM estimates to account for outliers and heteroskedasticity. The independent variables are the weights of stocks in various AI indices for 2022 and 2023, including the Equally Weighted Artificial Intelligence Index (AII), and the Time-Discounted Artificial Intelligence Indices with discount factors of 0.5 (TAII05) and 5 (TAII5X), as detailed in Section 3.3. The CAARs are calculated over a 61-trading-day event window following the ChatGPT launch. The estimation window spans 251 trading days prior to the event. The market model uses the S&P 500 Index (SPX) as the benchmark. Statistical significance is denoted by *, **, and *** for p-values less than 0.10, 0.05, and 0.01, respectively. Standard errors are reported in parentheses.

## 4.3. Analysis of Abnormal Returns in AI-classified Stocks

Building upon the analysis of CAARs across all NASDAQ-listed stocks (Section 4.2), we refine our focus to examine the subset of 582 stocks classified as AI-related based on our NLP-based methodology. Specifically, we aim to assess how effectively the Size-Weighted Artificial Intelligence Index (SAII) metrics for 2022 and 2023 explain the abnormal returns of these AI stocks following the launch of ChatGPT. Table 3 presents the regression results employing both OLS and robust MM-estimator techniques to evaluate this relationship. In these regressions, the independent variables are the index weights derived from the SAII for the years 2022 and 2023. These weights quantify the degree of AI-related disclosures by companies within the AI stock classification, effectively serving as proxies for the intensity of AI engagement as communicated through their 10-K filings.

The regression results indicate statistically significant positive coefficients for both the 2022 and 2023 SAII metrics, which reveal that a higher degree of AI-related disclosure is associated with greater CAARs among AI stocks following the ChatGPT launch. This finding supports the validity of our NLP-based classification method, demonstrating that the extent of AI engagement, as captured by our SAII metrics, has a meaningful and positive relationship with abnormal stock returns in the context of a significant AI-related market event. However, the absence of a comparative baseline or alternative AI classification methods limits our ability to fully assess the relative effectiveness and robustness of our NLP-based approach. Future research could enhance this analysis by incorporating additional variables or employing alternative AI engagement measures to provide a more comprehensive evaluation.

**Table 3. Regression results predicting CAARs for AI-classified Nasdaq stocks**

|  | (1)  Coef. (SE) | (2)  Coef. (SE) | (3)  Coef. (SE) | (4)  Coef. (SE) |
|---|---|---|---|---|
| SAII (2022) | 18.211**  (9.282) | 16.613**  (8.433) |  |  |
| SAII (2023) |  |  | 13.467*  (8.799) | 16.813**  (7.967) |
| Constant | 0.144***  (0.026) | 0.078***  (0.017) | 0.152***  (0.026) | 0.079***  (0.016) |
| R2 / Pseudo R2 | 0.007 | 0.008 | 0.003 | 0.009 |
| F / Chi2 | 0.0503 | 0.0491 | 0.170 | 0.035 |
| Model | OLS | MM | OLS | MM |

Note: Table 3 presents estimates from regressions predicting cumulative average abnormal returns (CAARs) for a subset of 582 NASDAQ-listed stocks classified as AI-related, following the ChatGPT launch, as specified in Equation (12). Models 1 and 3 provide ordinary least squares (OLS) estimates, while Models 2 and 4 offer robust

MM estimates. The independent variable is the weight of each stock in the Size-Weighted Artificial Intelligence Index (SAII) for 2022 and 2023, as detailed in Section 3.3. The CAARs are calculated over a 61-trading-day event window after the ChatGPT launch on November 30, 2022. The estimation window consists of 251 trading days prior to the event. The market model employs the S&P 500 Index (SPX) as the reference market. Statistical significance is indicated by *, **, and *** for p-values less than 0.10, 0.05, and 0.01, respectively. Standard errors are shown in parentheses.

## 4.4. Comparative Evaluation of AI Engagement in Indices and ETFs

To evaluate the effectiveness of our AI stock classification methodology at the index level, we compare the performance of our four AI indices with that of existing AI-themed ETFs and the Nasdaq Composite Index (IXIC) following the launch of ChatGPT. We conduct an event study using daily returns for our indices, the 12 AI-focused ETFs identified in Table 1, and the IXIC. The rationale is that if our NLP-based AI indices effectively capture AI-related market dynamics, they should exhibit significant positive abnormal returns in response to the ChatGPT launch, similar to existing AI ETFs designed to provide exposure to AI-related investments.

As reported in Table 4, all four of our AI indices experienced statistically significant positive cumulative abnormal returns (CARs) over the three-month event window, with an average CAR of 17.48%.[8] Specifically, the CARs for our indices ranged from 16.19% to 20.97%, with corresponding t-statistics all significant at the 5% level. This indicates that our indices effectively captured the positive market reaction associated with the ChatGPT launch. In comparison, only half of the existing AI ETFs (6 out of 12) exhibited significant positive abnormal returns over the same period, with an average CAR of 17.58% among these six ETFs. When considering all 12 AI ETFs, the average CAR was 13.00%; excluding the outlier ROBT—which had a negative CAR of -15.37%—the average CAR increased to 15.58%. Notably, three of the existing AI ETFs (IRBO, ARKQ, and BOTZ) reported the largest positive market reactions, with CARs of 22.80% and 21.40%.

Our four AI indices ranked fourth to seventh in terms of CAR magnitude among all indices and ETFs analyzed. This competitive positioning suggests that our objective, NLP-based AI stock classification method yields indices that are comparable to existing AI-themed ETFs in capturing market responses to significant AI-related events. The fact that our indices performed on par with or better than several existing ETFs provides empirical support for the efficacy of our methodology. These findings emphasize the potential of our objective, data-driven approach in constructing thematic investment indices. By effectively capturing the market's positive response to the ChatGPT launch, our AI indices demonstrate their ability to reflect investors' sentiment toward AI developments. This further validates our classification methodology and suggests that such an approach can offer a cost-effective and transparent alternative for thematic investing in the AI sector.

---

[8] Abnormal Return (AR) measures the difference between a stock's actual return and its expected return during the event window, indicating the market reaction to AI-related events. Cumulative Abnormal Return (CAR) aggregates ARs over multiple days to assess the overall impact, while Cumulative Average Abnormal Return (CAAR) averages CARs across all firms in the sample, offering a general view of how AI stocks responded to the event. These metrics collectively help quantify investor reactions to the launch of ChatGPT.

**Table 4. Event study results for AI indices and ETFs following ChatGPT launch**

| Ticker | Cumulative Abnormal Returns, CARs [0, 61] | Abnormal Returns, ARs [Per day] | t-statistic [significance] | z-statistic [significance] |
|---|---|---|---|---|
| Panel (a) AI Indices | | | | |
| AII | 16.35 | 0.27 | 2.50** | 2.27** |
| SAII | 20.97 | 0.34 | 2.61** | 2.32** |
| TAII05 | 16.19 | 0.27 | 2.50** | 2.19** |
| TAII5X | 16.40 | 0.27 | 2.43** | 2.21** |
| Panel (b) AI ETFs | | | | |
| IGPT | 8.94 | 0.15 | 1.34 | 1.19 |
| ROBO | 15.05 | 0.25 | 3.12*** | 2.83*** |
| ARKQ | 21.40 | 0.35 | 2.66*** | 2.61*** |
| BOTZ | 21.11 | 0.35 | 3.60*** | 2.93*** |
| AIQ | 14.70 | 0.24 | 2.08** | 2.38** |
| IRBO | 22.80 | 0.37 | 3.24*** | 3.06*** |
| QTUM | 10.47 | 0.17 | 2.18** | 1.73* |
| GOAI | 4.59 | 0.08 | 0.51 | 0.37 |
| WTAI | 16.67 | 0.27 | 1.13 | 1.07 |
| XAIX | 15.14 | 0.25 | 1.23 | 1.00 |
| ROBT | -15.37 | -0.25 | -1.11 | -1.34 |
| AIAG | 20.51 | 0.34 | 1.32 | 1.20 |
| Panel (c) Nasdaq Composite Index | | | | |
| IXIC | 7.69 | 0.13 | 2.46** | 2.38** |

Note: Table 4 summarizes the mean abnormal returns (AR) per day and cumulative abnormal returns (CARs) over the event window, expressed in percentages, for AI-themed indices and exchange-traded funds (ETFs) following the launch of OpenAI's ChatGPT on November 30, 2022. The estimation window includes 251 trading days (one year) prior to the event, and the event window spans 61 trading days (three months) after the launch. The market model uses the S&P 500 Index (SPX) as the reference market. The z-statistic corresponds to the nonparametric Wilcoxon signed-rank test. Statistical significance levels are denoted by *, **, and *** for p-values less than 0.10, 0.05, and 0.01, respectively.

## 4.5. Comprehensive Performance Assessment of AI Indices and ETFs

We extend our investigation to evaluate the overall performance of these indices relative to existing AI-focused ETFs and the broader market. This comprehensive assessment aims to examine both the short-term responsiveness and the long-term risk and return characteristics of our indices, providing a more holistic view of their effectiveness as investment vehicles in the AI sector. To achieve this, we employ a range of traditional financial performance metrics consistent with established practices in financial research (Häusler and Xia, 2022; Schröder, 2007; Zatlavi et al., 2014), allowing for a detailed evaluation over the period from June 27, 2019, to September 1, 2023. This time frame represents the longest common period with available market data for all indices and ETFs under consideration.

Central to our analysis is the average daily return percentage, $(Ret)$ which reflects the average daily percentage change in the value of an ETF or index. This metric provides a foundational comparison of the returns generated by different investment instruments. To assess risk, we examine the standard deviation $(SD)$ of returns, which measures the volatility or variability of returns over the specified period. A higher $SD$ indicates greater fluctuation in returns, suggesting higher investment risk. Additionally, we analyze skewness $(Skew)$ and kurtosis $(Kurt)$ to gain insights into the distribution of returns. Skewness measures the asymmetry of the return distribution. At the same time, kurtosis indicates the propensity for extreme values, both of which are crucial for understanding tail risks and the likelihood of significant deviations from the mean return, offering insights into the asymmetry and extremity of returns.

In order to contextualize the performance of our AI indices and ETFs relative to the broader market, we compute alpha and beta coefficients using the Nasdaq Composite Index (IXIC) as the benchmark market portfolio. Alpha $(\alpha)$ represents the intercept in a regression of the index, or ETF returns, against market returns, indicating the ability of the investment to generate excess returns independent of market movements. A positive alpha suggests outperformance relative to the market. Beta $(\beta)$ measures the sensitivity of the investment's returns to market movements, quantifying systematic risk. A beta greater than one indicates higher volatility compared to the market, while a beta less than one suggests lower volatility. We further evaluate the risk-adjusted performance using the Sharpe Ratio $(SR)$, calculated as:

$$SR = \frac{\bar{r} - r_i}{SD},\qquad(13)$$

where $r_i$ is the average return, and is the risk-free rate, operationalized using the 3-month U.S. Treasury bill yield. The Sharpe Ratio assesses the return earned per unit of total risk, facilitating comparisons across investments with different risk profiles (Sharpe, 1966). Complementing the Sharpe Ratio, we utilize the Sortino Ratio $(SoR)$, which focuses on downside risk by using the standard deviation of negative returns ($SD_{down}$) in the denominator, emphasizing the importance of penalizing only unfavorable volatility (Sortino and Price, 1994). This metric is particularly relevant for investors who are more concerned with negative deviations from expected returns.

$$SoR = \frac{\bar{r} - r_i}{SD_{down}},$$  (14)

To understand the potential for significant losses, we calculate the Maximum Drawdown ($MMD$), defined as:

$$MMD = \frac{P_{peak} - P_{trough}}{P_{peak}} \times 100.$$  (15)

where $P_{peak}$ is the highest value of the investment before a significant decline, and $P_{trough}$ is the lowest value following the peak. The $MMD$ provides insight into the worst-case loss scenario over the investment period, which is crucial for risk management and assessing an investor's exposure to potential downturns (Häusler and Xia, 2022).

Finally, we employ the Omega Ratio, $\omega(r_i)$, to assess performance relative to a minimum acceptable return, the risk-free rate ($r_i$), operationalized using 3-month US Treasury Bills:

$$\omega(r_i) = \frac{\int_{r_f}^{\infty} (1 - F(r)) dr}{\int_{-\infty}^{r_f} F(r) dr},$$  (16)

where $F(r)$ is the cumulative distribution function of returns. The Omega Ratio provides a comprehensive measure of the likelihood of achieving returns above a certain threshold, incorporating all moments of the return distribution and offering a more nuanced performance evaluation (Keating and Shadwick, 2002).

By applying spectrum of traditional and sophisticated financial metrics to our AI indices and comparing them with existing AI-focused ETFs and the IXIC, we aim to provide a detailed evaluation of their performance characteristics. This comprehensive assessment facilitates informed decision-making by investors and highlights the potential advantages of our objective, NLP-based approach in constructing thematic investment indices. It serves to reveal how our indices perform in various market conditions and whether they offer competitive returns relative to existing investment options in the rapidly evolving AI sector by comparing performance, risk, and return profiles.

Table 5 presents a comprehensive performance evaluation of the AI-focused indices and ETFs. The first four entries in Panel (a)—AII, SAII, TAII05, and TAII5X—are the indices we constructed based on NLP analysis of SEC filings. The next twelve entries in Panel (b) represent a diverse array of established AI-focused ETFs (cf. Table 1), while panel (c) offers a benchmark comparison with the Nasdaq Composite Index (IXIC). The performance metrics analyzed include average daily return percentage ($Ret$), standard deviation ($SD$), skewness ($Skew$), kurtosis ($Kurt$), alpha ($\alpha$), beta ($\beta$), Sharpe Ratio ($SR$), Sortino Ratio ($SoR$), maximum drawdown ($MMD$), and omega ratio $\omega(r_i)$. Collectively, these metrics provide a multifaceted view of each instrument's return characteristics, risk profile, volatility, and overall performance efficiency over the period from June 27, 2019, to September 1, 2023.

Examining the average daily returns presented in Table 5, our AI indices in Panel (a) exhibited a higher mean return ($ret$) of 0.076% compared to 0.056% for the existing AI ETFs in Panel (b). The Nasdaq Composite Index (IXIC) in Panel (c) reported an average daily return of 0.068%, falling between these two groups. This indicates that over the analyzed period, our AI indices outperformed both the market benchmark and the existing AI-themed ETFs in terms of raw returns. In terms of volatility, measured by the standard deviation ($SD$) of returns, both our AI indices (Panel a) and the AI ETFs (Panel b) displayed similar levels, with mean standard deviations of 1.8% and 1.7%, respectively. This suggests that the day-to-day fluctuations in returns were comparable across both groups, and neither exhibited significantly higher volatility than the other. The IXIC in Panel (c) had a slightly lower standard deviation of 1.7%, indicating marginally less volatility than our AI indices.

Analyzing skewness, both our AI indices in Panel (a) and the AI ETFs in Panel (b) exhibited negative skewness ($Skew$), with mean values of -0.44 and -0.42, respectively, reflecting a slight propensity for returns to deviate negatively from the mean. The IXIC's skewness in Panel (c) was -0.43, comparable to both groups. Regarding kurtosis ($Kurt$), higher values were observed for the AI ETFs (mean of 8.56) in Panel (b) compared to our AI indices (mean of 7.30) in Panel (a), while the IXIC in Panel (c) exhibited the highest kurtosis at 9.14. Elevated kurtosis indicates a greater likelihood of extreme returns (both positive and negative), suggesting that the AI ETFs in Panel (b) were more susceptible to such events than our AI indices in Panel (a).

Evaluating risk-adjusted returns using the Sharpe Ratio and Sortino Ratio, our AI indices (Panel a) demonstrated superior performance, with mean ($SR$) and ($SoR$) values of 0.039 and 0.037, respectively, compared to 0.029 and 0.028 for the AI ETFs (Panel b). This indicates that, per unit of risk, our AI indices provided, on average, higher returns than the existing AI ETFs. QTUM achieved the highest individual Sharpe Ratio (0.046) and Sortino Ratio (0.044) amongst all existing AI ETFs in Panel (b). Analyzing beta ($\beta$) coefficients, which measure sensitivity to market movements, our AI indices in Panel (a) had a higher mean beta of 0.961 compared to 0.623 for the AI ETFs in Panel (b). This suggests that our indices were more closely correlated with the market benchmark (IXIC) and tended to move more in line with overall market trends. The higher beta of our indices (Panel a) is consistent with their construction from NASDAQ-listed stocks. In contrast, the AI ETFs (Panel b) may include stocks from other exchanges, potentially reducing their correlation with the IXIC. Among the AI ETFs, WTAI had a notably low beta of 0.206, indicating a lower correlation with the market and potential diversification benefits within a portfolio context.

In terms of maximum drawdown ($MMD$), our AI indices in Panel (a) experienced a slightly lower mean $MMD$ of 4.65% compared to 4.92% for the AI ETFs in Panel (b), suggesting a marginally smaller historical peak-to-trough loss. The IXIC in Panel (c) had an even lower MMD of 4.12%, indicating that the broader market experienced smaller drawdowns over the period. Lastly, the omega ratio, which assesses performance relative to a minimum required return, was equal on average for both our AI indices (Panel a) and the AI ETFs (Panel b) at 1.29. This indicates that, overall, both groups offered similar probabilities of achieving returns above the risk-free rate. The IXIC, however, had a higher omega ratio of 1.40, suggesting a better return per unit of risk compared to both AI-focused groups. On an individual level, among the AI ETFs, XAIX and GOAI both achieved the highest omega ratios of 1.44, outperforming even the IXIC in this measure.

Overall, our AI indices constructed through an NLP-based methodology outperformed existing AI ETFs by delivering higher average daily returns and demonstrated superior risk-adjusted performance, as evidenced by higher Sharpe and Sortino Ratios. Importantly, this enhanced performance did not come at the expense of increased volatility, as indicated by the comparable standard deviations between our indices and the ETFs. The higher beta values of our indices reflect a stronger alignment with overall market movements, which may be advantageous for investors seeking exposure to the AI sector that closely tracks market trends. While certain AI ETFs exhibited unique characteristics—such as WTAI's low beta offering diversification benefits or QTUM and XAIX's strong individual performance metrics—the overall findings suggest that our objective AI indices more effectively capture the innovative essence of the AI sector while maintaining a risk profile comparable to existing investment options. Consequently, our indices provide a compelling alternative for investors interested in thematic AI investing, combining robust returns with acceptable risk levels in line with the broader market as represented by the IXIC, thereby highlighting the potential advantages of our data-driven approach in constructing thematic investment indices.

**Table 5. Performance evaluation of AI indices and ETFs compared to the Nasdaq composite index**

| Ticker | Ret | SD | Skew | Kurt | Alpha | Beta | SR | SoR | MMD | $\omega(r_i)$ |
|---|---|---|---|---|---|---|---|---|---|---|
| Panel (a) AI Indices | | | | | | | | | | |
| AII | 0.075 | 0.018 | -0.49 | 7.60 | 0.00011 | 0.945 | 0.038 | 0.036 | 4.70 | 1.34 |
| SAII | 0.089 | 0.019 | -0.36 | 6.91 | 0.00021 | 0.994 | 0.043 | 0.042 | 4.52 | 1.23 |
| TAII05 | 0.074 | 0.018 | -0.48 | 7.47 | 0.00009 | 0.945 | 0.038 | 0.036 | 4.74 | 1.33 |
| TAII5X | 0.072 | 0.018 | -0.44 | 7.20 | 0.00006 | 0.960 | 0.036 | 0.035 | 4.63 | 1.24 |
| Mean | 0.076 | 0.018 | -0.44 | 7.30 | 0.00011 | 0.961 | 0.039 | 0.037 | 4.65 | 1.29 |
| Panel (b) AI ETFs | | | | | | | | | | |
| IGPT | 0.026 | 0.018 | -0.42 | 5.98 | 0.00040 | 0.970 | 0.011 | 0.010 | 3.33 | 1.08 |

| | Ret | SD | Skew | Kurt | $\alpha$ | $\beta$ | SR | SoR | MMD | $\omega(r_i)$ |
|---|---|---|---|---|---|---|---|---|---|---|
| ROBO | 0.047 | 0.017 | -0.28 | 8.87 | -0.00016 | 0.920 | 0.024 | 0.024 | 3.84 | 1.12 |
| ARKQ | 0.074 | 0.021 | -0.17 | 4.90 | -0.00003 | 1.130 | 0.032 | 0.031 | 3.39 | 1.20 |
| BOTZ | 0.044 | 0.019 | -0.19 | 9.06 | -0.00026 | 1.013 | 0.020 | 0.019 | 3.84 | 1.13 |
| AIQ | 0.070 | 0.017 | -0.23 | 5.88 | 0.00003 | 0.975 | 0.037 | 0.035 | 3.50 | 1.38 |
| IRBO | 0.043 | 0.018 | -0.26 | 6.53 | -0.00024 | 0.977 | 0.021 | 0.020 | 3.90 | 1.31 |
| QTUM | 0.087 | 0.018 | -0.40 | 8.07 | 0.00020 | 0.980 | 0.046 | 0.044 | 3.17 | 1.33 |
| GOAI | 0.055 | 0.013 | -0.45 | 10.40 | 0.00052 | 0.042 | 0.038 | 0.035 | 5.32 | 1.44 |
| WTAI | 0.076 | 0.018 | -0.18 | 7.34 | 0.00062 | 0.206 | 0.040 | 0.038 | 3.89 | 1.26 |
| XAIX | 0.066 | 0.014 | -0.30 | 7.49 | 0.00051 | 0.220 | 0.043 | 0.041 | 3.98 | 1.44 |
| ROBT | 0.037 | 0.017 | -0.76 | 11.07 | 0.00039 | -0.031 | 0.018 | 0.017 | 4.16 | 1.40 |
| AIAG | 0.041 | 0.019 | -1.37 | 17.13 | 0.00036 | 0.070 | 0.018 | 0.017 | 16.66 | 1.36 |
| Mean | 0.056 | 0.017 | -0.42 | 8.56 | 0.00018 | 0.623 | 0.029 | 0.028 | 4.92 | 1.29 |
| Panel (c) Nasdaq Composite Index | | | | | | | | | | |
| IXIC | 0.068 | 0.017 | -0.43 | 9.14 | - | - | 0.037 | 0.035 | 4.12 | 1.40 |

Note: Table 5 presents performance metrics for AI-focused indices and ETFs, along with the Nasdaq Composite Index (IXIC), over the period from June 27, 2019, to September 1, 2023 (N = 1,054 trading days), representing the longest common time frame with available market data for all instruments. Panel (a) includes the four AI indices we constructed based on NLP analysis of SEC filings, while Panel (b) comprises twelve existing AI-focused ETFs identified in Table 1. Panel (c) provides the performance metrics for the IXIC as a market benchmark. Metrics reported are daily return percentage ($Ret$), standard deviation of returns ($SD$), skewness ($Skew$), kurtosis ($Kurt$), alpha ($\alpha$), and beta ($\beta$) calculated with respect to the IXIC. The Sharpe Ratio ($SR$) and Sortino Ratio ($SoR$) are computed using the 3-month U.S. Treasury bill yield as the risk-free rate ($r_i$). Additionally, the maximum drawdown ($MMD$) and the omega ratio $\omega(r_i)$ are provided.

## 5. Implications

This study's primary contribution lies in the development and validation of a novel measure for the presence of AI-related disclosures in stocks, representing an advancement in the realm of thematic investing. By leveraging natural language processing (NLP) techniques to analyze corporate filings, our measure offers a fresh perspective on the centrality of AI within publicly traded companies. Importantly, our approach to stock classification and valuation is not confined to AI; it is theoretically applicable to any technology or thematic focus, given its reliance on predefined keywords and publicly available disclosures. The validation of our AI measure, particularly against the backdrop of the launch of OpenAI's ChatGPT, supports its effectiveness and highlights its relevance in today's technology-driven investment landscape.

A particularly striking finding is the competitiveness of our AI indices when compared to established AI-themed exchange-traded funds (ETFs). In several instances, our objectively constructed indices matched or even exceeded the performance metrics of traditional AI ETFs, a noteworthy achievement considering the substantial expense ratios often associated with such investment products. This observation suggests that a straightforward, data-driven approach, as exemplified by our AI stock measure, can yield performance outcomes that are on par with or surpass those of more complex and costly investment vehicles. This finding is especially pertinent in the current investment environment, where cost efficiency and effective asset selection are of paramount importance.

## 5.1.  Implications for Theory

The findings of this study have meaningful theoretical implications in the fields of finance and innovation, particularly concerning the role of artificial intelligence (AI) in contemporary markets. Firstly, our research contributes to the literature on market reactions to technological advancements by providing a tangible metric that quantifies the market's perception of AI's value within stocks. By developing an AI measure based on natural language processing (NLP) analysis of corporate filings, we enhance existing theories on investor behavior in response to technological innovations (Aharon et al., 2022; Cahill et al., 2020). Consistent with the Efficient Market Hypothesis (Fama, 1970), our results suggest that markets can rapidly assimilate and respond to information regarding a company's engagement with AI technologies.

Secondly, this study enriches the discourse on thematic investing (Somefun et al., 2023). Our findings— particularly the competitive performance of AI indices relative to established AI-themed exchange-traded funds (ETFs)—support the notion that thematic investments, when well-constructed and data-driven, can be both effective and cost-efficient. This insight adds depth to the existing literature on thematic investments and ETFs (Blitz, 2021b; Dheeriya and Malladi, 2019; Methling and von Nitzsch, 2019), suggesting that objective, data-driven methodologies can enhance the effectiveness of thematic investment strategies.

Thirdly, our research contributes to the strategic management literature by more closely linking AI integration in corporate strategies with firm valuation and competitive advantage (Kar et al., 2021; Keding, 2021). By demonstrating how AI integration is perceived and valued by the market, we expand on theories that explore psychological factors influencing investment decisions in behavioral finance (Bajo et al., 2023; Barber and Odean, 2013).

Fourthly, our results prompt a re-examination of the role of corporate disclosure in an era increasingly dominated by AI and technology. They suggest the need for new frameworks to assess the quality and impact of such disclosures, particularly in technology-driven sectors. The NLP-based measure introduced in this study exemplifies how qualitative information can be transformed into a quantifiable metric with significant market relevance. This resonates with the current shift towards more sophisticated data analysis methods in financial research and emphasizes the importance of technological literacy in financial market analysis.

Finally, our methodology can be readily extended to other emerging technological themes, such as blockchain, the metaverse, or environmental, social, and governance (ESG) topics. Offering comparative insights contributes to a broader understanding of how different technologies are valued in financial markets and provides a benchmark for existing indices and ETFs. This could inspire economic theories regarding the broader impact of technologies like AI on market structures, competition, and economic growth, particularly as AI continues to permeate various sectors.

## 5.2.   Implications for Practice

Our study offers valuable insights for a broad spectrum of stakeholders in the financial sector, with implications ranging from investment strategy formulation to risk management. Firstly, our study provides valuable insights for investors, particularly those interested in thematic investing. The marked outperformance of AI stocks following the launch of ChatGPT highlights the notable potential of AI as an investment theme. This remarkable performance reflects AI's escalating prominence in the investment landscape and offers new opportunities for portfolio diversification and growth. Investors can leverage the insights from our AI stock measure to identify companies deeply integrated with AI technologies, potentially leading to enhanced returns and aligning their portfolios with cutting-edge technological advancements.

Secondly, for fund managers and index providers, the findings present a compelling case for adopting more data-driven and transparent methodologies in constructing thematic indices. The competitive performance of the AI indices developed in this research—rooted in objective analyses using publicly available corporate data—suggests that such methods can yield outcomes comparable to, or even surpass, those achieved by existing thematic ETFs. This approach could enhance competition among index providers, potentially resulting in reduced costs for ETFs and investors (An et al., 2023). Moreover, the methodology employed in this study could revolutionize thematic index construction, serving as a blueprint for developing similar measures across other thematic areas. By harnessing technologies like NLP, fund managers and analysts can extract actionable insights from corporate filings, enabling the formulation of more nuanced and accurate investment strategies.

Thirdly, regulators and policymakers might find value in these findings, as they highlight the importance of transparent corporate disclosures in promoting market efficiency. Encouraging companies to provide detailed information about their engagement with emerging technologies like AI can facilitate more informed investment decisions and enhance investor protection. Additionally, the adoption of objective, data-driven approaches align with the broader trend toward increased transparency and accountability in financial markets. This could lead to the development of regulatory frameworks that standardize AI-related disclosures, further improving the quality of information available to investors.

Fourthly, our study has implications for risk management practices within financial institutions. The rapidly evolving nature of AI technologies introduces unique risks and uncertainties. By providing a quantifiable measure of AI engagement, our methodology aids investors and risk managers in assessing the exposure of their portfolios to AI-related risks and opportunities. This enables more effective risk assessment and the development of strategies to mitigate potential adverse impacts associated with technological disruptions.

Finally, our research emphasizes the practical utility of applying advanced analytical techniques to publicly available data, offering a cost-effective and transparent alternative for financial technology investors. The demonstrated feasibility and success of a straightforward, data-driven approach to thematic investing suggest that such methods hold the potential to outperform traditional, more costly strategies. This highlights the value of technological literacy and innovation in financial analysis, encouraging stakeholders to embrace data analytics and NLP techniques in investment decision-making processes.

## 5.3. Limitations and Future Research

This study, while providing valuable insights, is subject to several limitations that offer avenues for future research and refinement. Firstly, our reliance on publicly available corporate disclosures, specifically 10-K filings, means that our quantitative analysis may not capture all dimensions of a company's engagement with AI. A closer examination of these filings reveals that the term "AI" is used in varied contexts. Some companies detail how they have integrated AI into their business operations, while others merely mention competitors' use of AI or express intentions to explore AI in the future. Future research could enhance the quality of these metrics by incorporating qualitative assessments or employing more advanced natural language processing techniques to differentiate the context and depth of AI mentions (Frankel et al., 2022).

Secondly, the utilization of a single data source presents limitations. Companies could theoretically manipulate their AI engagement metrics by frequently mentioning AI without substantive involvement, a phenomenon known as "AI washing." Indeed, in March 2024, the SEC charged two investment advisers for disclosing misleading statements about AI utilization (SEC, 2024), indicating that such manipulation is a genuine concern. To mitigate this issue, future studies should consider incorporating additional data sources, such as research and development (R&D) expenditures (Mishra et al., 2022), technology sentiment analyses (Bonaparte, 2023), or other AI-related keywords like "machine learning." This multi-faceted approach could develop a more robust and comprehensive measure of AI engagement, reducing the potential for manipulation and improving the reliability of the classification.

Thirdly, the market reaction to the launch of ChatGPT-3 launch supports our AI stock classification; this event-based approach provides valuable insights into immediate market perceptions and responses, but it may not fully capture broader market dynamics or external factors influencing stock performance. Investor sentiment can be heavily influenced by short-term news cycles and hype, potentially skewing the perceived value of AI stocks. Therefore, future research should complement this event study with more holistic analyses that consider a range of market influences, longer time horizons, and multiple events to provide a more comprehensive understanding of AI stock performance and classification.

Fourthly, the focus on NASDAQ-listed firms presents a geographical limitation. Since the NASDAQ predominantly features U.S.-based technology companies, our analysis may overlook AI investment opportunities in other global markets, including emerging economies where AI technology is increasingly significant. This narrow focus restricts our understanding of the global AI investment landscape. Future research should expand the scope to include diverse markets and assess different types of corporate disclosures beyond 10-K filings, such as annual reports, earnings calls, and international regulatory filings, to capture a more accurate and comprehensive picture of global AI engagement.

Lastly, while our measure effectively identifies AI-centric stocks, it represents only a starting point in the complex task of evaluating AI's role and impact within companies and financial instruments. Our methodology is primarily quantitative, and incorporating qualitative assessments and more sophisticated metrics could enhance the precision of the AI engagement measure. For example, adapting methods from environmental, social, and governance (ESG) research—such as developing an AI-specific dictionary similar to the ESG dictionary by Baier et al., (2020) —could provide deeper insights into the quality and depth of AI integration within firms. Incorporating relevant metrics like research and development (R&D) expenses (Mishra et al., 2022), sentiment analysis (Bonaparte, 2023), or additional AI-related keywords such as "machine learning" could also refine the measure. We propose that the quality of the AI measure can be further enhanced by conducting a broader and more in-depth examination of how AI is integrated and operationalized within the business models and strategies of these companies. Nevertheless, given the promising outcomes produced by our current approach, it is reasonable to infer that a more comprehensive and nuanced model could yield highly competitive results.

# 6. Conclusion

This study introduces an objective, data-driven methodology for classifying AI stocks based on publicly available corporate disclosures. By leveraging natural language processing techniques to analyze annual 10-K filings from 3,395 NASDAQ-listed firms between 2011 and 2023, we quantified each company's engagement with AI technologies. Our approach involved creating both a binary indicator of AI mention and a weighted AI score that reflects the frequency and context of AI-related terms in these disclosures. Using these AI engagement metrics, we constructed four AI stock indices: the Equally Weighted AI Index (AII), the Size-Weighted AI Index (SAII), and two Time-Discounted AI Indices (TAII05 and TAII5X). These indices differ in how they weigh companies based on their AI involvement and account for historical disclosures, offering alternative perspectives on AI investment opportunities. We then compared the performance of these indices against 12 existing AI-themed ETFs, analyzing their risk-return profiles, market reactions to the ChatGPT launch, and overall investment performance.

Our validation through an event study centered on the launch of ChatGPT confirmed that companies with higher degrees of AI engagement exhibited more significant positive abnormal returns. This finding demonstrates that our objective measures effectively capture market perceptions of AI involvement. Regression analyses further supported the validity of our metrics, revealing a significant predictive power of our AI measures on the cumulative average abnormal returns observed. Comparative analysis showed that our objectively constructed AI indices perform on par with, or even surpass, existing AI-themed ETFs in terms of market response and investment performance. This suggests that our methodology offers a reliable and market-responsive approach to creating AI stock indices, providing investors with a viable alternative or complement to existing AI-themed investment products. Such an approach could enhance competition among index providers and potentially reduce ETF expense ratios, benefiting investors seeking exposure to the AI sector.

Our study contributes to the literature on financial markets and corporate disclosures, particularly focusing on AI technology. By introducing an objective, data-driven approach to classifying AI stocks based on company filings, we address a notable gap in current research. Building upon foundational work on how corporate information influences stock valuations, our results enhance the understanding of market reactions to technology disclosures in corporate filings. Specifically, by analyzing the market reaction to the launch of OpenAI's ChatGPT as a validation measure for AI-focused stocks, we contribute to research on technology launches and their impact on financial markets. Overall, our study presents a robust framework for objectively identifying and evaluating AI-focused investments. By utilizing NLP-based analysis of corporate disclosures, we create data-driven AI engagement metrics that inform the construction of competitive AI stock indices. Our methodology addresses the opacity and subjectivity found in existing AI-themed ETFs and offers empirical support through market reaction analyses. As AI continues to evolve and impact various sectors, our methodologies and findings remain relevant for understanding and capitalizing on AI-related investment opportunities, benefiting both academia and the investment industry.

## Statements

During the preparation of this work, we utilized ChatGPT-4 and Grammarly to enhance language and readability. After employing these tools, the content was thoroughly reviewed and edited as necessary, and the authors took full responsibility for the final version of the publication.

**Appendix**

**Figure A.1. Mean daily returns and expense ratios of AI-themed ETFs**

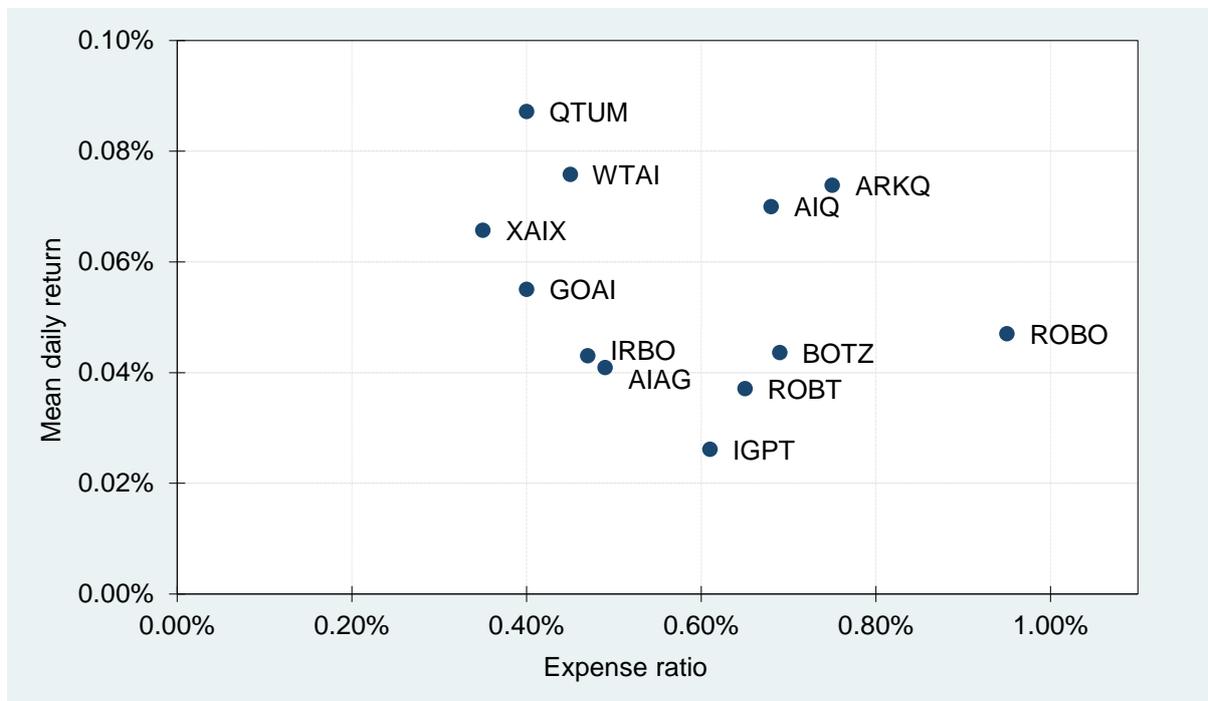

Note: This figure presents a scatter plot comparing the mean daily returns and expense ratios of AI-themed ETFs listed in Table 1. Each data point represents an individual ETF, plotted based on its expense ratio and corresponding mean daily return over the period from June 27, 2019, to September 1, 2023 (N = 1,054).